\documentclass[a4paper,11pt,oneside]{article}

\usepackage[english]{babel}
\usepackage[T1]{fontenc}
\usepackage[utf8]{inputenc}

\usepackage[pdftex,unicode]{hyperref}
\hypersetup{pdftitle=The Role of a Nation's Culture in the Country's Governance: Stochastic Frontier Analysis}
\hypersetup{pdfauthor=Vladimír Holý and Tomáš Evan}

\usepackage{color}
\definecolor{myred}{rgb}{0.60,0.30,0.00}
\hypersetup{colorlinks=true, linkcolor=myred, anchorcolor=myred, citecolor=myred, filecolor=myred, urlcolor=myred}

\usepackage[margin=60pt]{geometry}
\usepackage{amsmath}
\usepackage{amssymb}
\usepackage{bm}

\usepackage[group-minimum-digits=3]{siunitx}

\usepackage{enumitem}

\usepackage{graphicx}
\usepackage{float}
\usepackage{hhline}
\usepackage{multirow}
\usepackage{rotating}
\usepackage{booktabs}

\usepackage[authoryear]{natbib}

\begin{document}

\begin{center}
{\Large \bfseries The Role of a Nation's Culture in the Country's Governance: \\ Stochastic Frontier Analysis}
\end{center}

\begin{center}
{\bfseries Vladimír Holý} \\
Prague University of Economics and Business \\
Winston Churchill Square 1938/4, 130 67 Prague 3, Czech Republic \\
\href{mailto:vladimir.holy@vse.cz}{vladimir.holy@vse.cz}
\end{center}

\begin{center}
{\bfseries Tomáš Evan} \\
Czech Technical University in Prague \\
Thákurova 2077/7, 160 00 Prague 6, Czech Republic \\
\href{mailto:tomas.evan@fit.cvut.cz}{tomas.evan@fit.cvut.cz}
\end{center}

\begin{center}
{\itshape \today}
\end{center}

\noindent
\textbf{Abstract:}
What role does culture play in determining institutions in a country? This paper argues that the establishment of institutions is a process originating predominantly in a nation's culture and tries to discern the role of a cultural background in the governance of countries. We use the six Hofstede's Cultural Dimensions and the six Worldwide Governance Indicators to test the strength of the relationship on 94 countries between 1996 and 2019. We find that the strongest cultural characteristics are Power Distance with negative effect on governance and Long-Term Orientation with positive effect. We also determine how well countries transform their cultural characteristics into institutions using stochastic frontier analysis.
\\

\noindent
\textbf{Keywords:} Hofstede's Cultural Dimensions, Worldwide Governance Indicators, Technical Efficiency, Stochastic Frontier Analysis.
\\

\noindent
\textbf{JEL Codes:} C23, C44, O15, O43.
\\

\section{Introduction}
\label{sec:intro}

In recent years, the growing body of literature claims a strong relationship between culture as defined and measured by \cite{Hofstede2010} and institutions as systematically described i.a.\ by \cite{North1990}, \cite{Olson1996, Olson1998} and a large variety of socio-economic phenomena. Culture and economic institutions are both independently important for economic development but their impact is stronger when combined \citep{Williamson2011}. This combined effect is at least of the same magnitude as variables of the standard production function \citep{Evan2021}. The relationship between culture and institutions remains, however, notoriously complex and thus difficult to measure. In this paper we have decided to discern the role of a cultural background in governance of a country. Specifically, we have two goals:

\begin{enumerate}
\item To examine the strength of the relationship between cultural characteristics of countries and the quality of their institutions.
\item To determine how well individual countries transform their cultural characteristics into institutions.
\end{enumerate}

In line with \cite{Williamson2011}, we consider that the establishment of institutions is a process originating in a nation's culture. Holders of a culture conducive to development may choose to formalize informal institutions into formal ones thus promoting this development. To this end, we employ stochastic frontier analysis of \cite{Aigner1977} and \cite{Meeusen1977}, which studies how efficiently a producer can transform inputs into outputs. As inputs, we use Hofstede's Cultural Dimensions \citep{Hofstede2010} measuring the characteristics of a culture. As outputs, we use Worldwide Governance Indicators \citep{Kaufmann2011} measuring the quality of institutions. Furthermore, to capture the economic environment, we use the gross domestic product (GDP) per capita. This framework allows us to achieve both of our research goals.

The rest of the paper is structured as follows. In Section \ref{sec:theory}, we lay out the theoretical foundations and review the related economic literature. In Section \ref{sec:ind}, we describe the cultural dimensions and the governance indicators. In Section \ref{sec:model}, we specify the stochastic frontier model we use. In Section \ref{sec:res}, we discuss the results of the efficiency analysis. We conclude the paper in Section \ref{sec:con}.

\section{Culture, Institutions, and Economic Development}
\label{sec:theory}

The idea of a significant cultural influence over institutions and governance is over a hundred years old \citep{Weber1905} but was strongly criticised by both Marxists \citep{Grossman2006} and by mainstream economists and remains a hotly debated issue to this day \citep{Blum2001, Becker2009, Cantoni2015, Kersting2020}. Mainstream economics has not readily even accepted institutions as a source of economic growth and development, innovation or quality of government, as the criticisms of the separate school of New Institutional Economics attests (\citealp{Coase1960, Coase1998, North1968, North1990, Olson1998, Hall1999}, i.a.). The idea has thus only very slowly made its appearance in economics before growing exponentially in recent years. There are papers using statistical analysis linking culture to innovation \citep{Zien1997, Tellis2009, Williams2010}, population growth \citep{Shennan2001, Yacout2015, Kumar2019}, environmental issues \citep{Peng2009, Nagy2018, Dangelico2020}, tax systems and collection \citep{Alm2006, Koenig2012, Cabelkova2013}, corruption \citep{Huber2001, Yeganeh2014}, software piracy \citep{Husted2000, Simmons2002}, and terrorism \citep{Meierrieks2013, Kluch2017} as well as a vast array of other systems or institutions. The first meta-analyses have appeared \citep{Taras2010, Buschgens2013} as there is enough data to attempt more complex studies. It seems the pendulum has swung from the cautious Boettke: “we cannot assume away cultural influences as economists have often done” \citep[p.\ 436]{Boettke2009} to the more radical Landes: “Max Weber had it right. If we learn anything from the history of economic development, it is that culture makes almost all the difference” \citep[p.\ 2]{Landes2000}. 

The idea that culture has a significant impact on the reality surrounding us either directly or indirectly through institutions in a variety of situations can be accepted. There is agreement on the primacy of human marketable capital, or culture as defined by \cite{Olson1996} for both the improvement of institutions (e.g.\ democratization) and economic growth. There is also overwhelming evidence for a causal link between particular economic institutions, most notably property rights and economic freedom, and economic development (for a recent literature review and summary of evidence, please consult i.e.\ \citealp{Feld2008, Acemoglu2010, Czegledi2014, Wanjuu2017}). To establish a definite causal link between institutions and economic growth has proven difficult. It means among other things understanding the technology of the transmission of institutional quality to economic growth and development. This includes three challenges as described by \cite{Docquier2014}, i.e.\ (i) disentangling  the causal effects and reversing the causal effects, (ii) accounting for unobserved shocks affecting both institutions and growth, and (iii) capturing the lag structure of the relationship. \cite{Acemoglu2005} claims political institutions as a fundamental cause of long-run economic growth. According to \cite{Acemoglu2005}, the knowledge of political institutions and the distribution of resources, or de facto political power, are sufficient to determine all the other variables in the system.

Some authors suggested that current measurement strategies have conceptual flaws and Human capital rather than political institutions have a causal effect on economic growth \citep{Djankov2003, Glaeser2004} also hinting at potentially significant reverse causal effects: “institutional outcomes also get better as the society grows richer, because institutional opportunities improve” \citep[p.\ 298]{Glaeser2004}. The standard institutionalists' approach that civil rights and democratic institutions cause development has been rather weakened in recent years as several studies using Granger regressions have found no evidence of this causality \citep{Paldam2012, Murtin2014}. Some authors returned to Lipset's hypothesis \citep{Lipset1959, Barro1997} reversing that causation and suggesting that development leads to democracy and civil rights and the issue remains unresolved \citep{Acemoglu2014, Jung2014, Czegledi2015}.

While we do not fully understand the relationship between institutions and economic growth or development it is more difficult still to determine the relationship between culture and institutions, or which one is more powerful in explaining the development in societies. A point in case is “Long-Term Persistence”, a paper by \cite{Guiso2016} in which they argue that Italian cities experiencing self-governance in the Middle Ages had a higher level of civic capital than other Italian cities. They offer three hypotheses with different causalities between culture and institutions (participation in public life in communes teaches people to cooperate and it is not forgotten; past democratic institutions change levels of trust and fairness in societies; historical events lead to changes in socialisation). 

Apart from complexity and long time periods for change, the issue of complementarities arises between culture and institutions hindering definitions and identification of channels of causality \citep{Alesina2015}\footnote{Please also refer to \cite{Alesina2015} for a literature review of culture and institutions.}. The complementarity hypothesis suggests the strength of the impact of culture and institutions lies in their combination. The impact of a high trust culture complemented by a high level of rule of law enforcement on the business environment would be a good example. Both culture and institutions would be significant in regression analysis. \cite{Williamson2011} shed some light on the issue by testing it. They claim it is the substitution effect between culture and institutions which is more important. They describe this important mechanism: “a culture conducive to economic growth may choose to formalize the informal institutions into institutions associated with economic freedom” (p.\ 316). Once the formal rules (institutions) are credible “the informal norms and mechanisms once relied upon for economic interaction and exchange, such as trust networks, may be rendered much less important” (p.\ 316). Culture is important when institutions promoting economic freedom and thus growth are absent but diminishing in significance when those institutions are established. This implies the different relative importance of culture and institutions depending on the level of economic development of particular countries. Less developed countries should ceteris paribus have a stronger influence of culture on institutions than more developed ones.

\section{Cultural Dimensions and Governance Indicators}
\label{sec:ind}

The term culture in its broad sense includes both civic culture (formal, institutions) and personal culture (informal) (\citealp{North1990, Olson1996}, i.a.). In this text, we use the narrow definition referring only to personal culture as we want to discern the difference between the two. To measure culture, we use \emph{Hofstede's Cultural Dimensions}. In our efficiency analysis, these dimensions act as input variables. \cite{Hofstede2010} define the six cultural dimensions in the following way:
\begin{enumerate}
\item \emph{Power Distance (PDI)} is the extent to which the less powerful members of institutions and organizations within a country expect and accept that power is distributed unequally.
\item \emph{Individualism (IDV)} pertains to societies in which the ties between individuals are loose: everyone is expected to look after him- or herself and his or her immediate family. Collectivism as its opposite pertains to societies in which people from birth onward are integrated into strong, cohesive in-groups, which throughout people’s lifetime continue to protect them in exchange for unquestioning loyalty.
\item \emph{Masculinity (MAS)} refers to society where emotional gender roles are clearly distinct: men are supposed to be assertive, tough, and focused on material success, whereas women are supposed to be more modest, tender, and concerned with the quality of life. A society is called feminine when emotional gender roles overlap: both men and women are supposed to be modest, tender, and concerned with the quality of life.
\item \emph{Uncertainty Avoidance (UAI)} is the extent to which the members of a culture feel threatened by ambiguous or unknown situations.
\item \emph{Long-Term Orientation (LTO)} stands for the fostering of virtues oriented toward future rewards -- in particular, perseverance and thrift. Its opposite pole, short-term orientation, stands for the fostering of virtues related to the past and present -- in particular, respect for tradition, preservation of “face”, and fulfilling social obligations.
\item \emph{Indulgence (IVR)} stands for a tendency to allow relatively free gratification of basic and natural human desires related to enjoying life and having fun. Its opposite pole, restraint, reflects a conviction that such gratification needs to be curbed and regulated by strict social norms.
\end{enumerate}
We standardize the variables to lie between 0 and 1. Note that as culture is viewed as unchangeable, the variables are static. The source of the data is \cite{Hofstede2020} with the methodology described in \cite{Hofstede2010}. Alternatively, the personal culture could be measured by Schwartz's Theory of Basic Human Values and Tabellini's Indicators of Individual Values and Beliefs. The theory of Hofstede's cultural dimensions, however, remains the most popular tool. For a survey of the literature based on Hofstede's Cultural Dimensions, we refer to \cite{Kirkman2006} and \cite{Beugelsdijk2017}.

To define civic culture or institutions we use the \emph{Worldwide Governance Indicators} consisting of six indicators for over 200 countries and territories over the period 1996--2019. In our efficiency analysis, these indicators play the role of output variables. \cite{Kaufmann2011} define the six governance indicators in the following way:
\begin{enumerate}
\item \emph{Voice and Accountability (VA)} captures perceptions of the extent to which a country's citizens are able to participate in selecting their government, as well as freedom of expression, freedom of association, and a free media.  
\item \emph{Political Stability and Absence of Violence/Terrorism (PV)} captures perceptions of the likelihood that the government will be destabilized or overthrown by unconstitutional or violent means, including politically‐motivated violence and terrorism.
\item \emph{Government Effectiveness (GE)} captures perceptions of the quality of public services, the quality of the civil service and the degree of its independence from political pressures, the quality of policy formulation and implementation, and the credibility of the government's commitment to such policies.
\item \emph{Regulatory Quality (RQ)} captures perceptions of the ability of the government to formulate and implement sound policies and regulations that permit and promote private sector development.
\item \emph{Rule of Law (RL)} captures perceptions of the extent to which agents have confidence in and abide by the rules of society, and in particular the quality of contract enforcement, property rights, the police, and the courts, as well as the likelihood of crime and violence.
\item \emph{Control of Corruption (CC)} captures perceptions of the extent to which public power is exercised for private gain, including both petty and grand forms of corruption, as well as "capture" of the state by elites and private interests.
\end{enumerate}
The variables are standardized to have zero mean and unit standard deviation. Higher values correspond to better governance. The source of the data is \cite{WGI2020} with the methodology described in \cite{Kaufmann2011}. 

Finally, we consider the gross domestic product (GDP) per capita to be a control variable representing the economic environment. For the importance of inclusion of the operating environment in an efficiency analysis, see e.g.\ \cite{Holy2020a}. To remove the trend in time but to keep differences in levels between countries, we define our \emph{GDP Level} variable as the difference between the logarithm of GDP per capita in USD current prices and the logarithm of the mean GDP per capita in each year. Note that suitable specification of the GDP variable is of great importance as it can influence results of an efficiency analysis (see e.g.\ \citealp{Holy2018e}). The source of the data is \cite{EconomicOutlook2020}.

\section{Stochastic Frontier Model}
\label{sec:model}

To capture dependence between the inputs (Hofstede's Cultural Dimensions) and the outputs (the Worldwide Governance Indicators) and assess efficiency of the individual countries, we utilize the framework of stochastic frontier analysis of \cite{Aigner1977} and \cite{Meeusen1977}. We also include GDP Level as a control variable. We consider a separate stochastic frontier model for each output variable but employ the same structure for all six models.  For a textbook treatment of stochastic frontier analysis, we refer to \cite{Kumbhakar2000}, \cite{Coelli2005}, \cite{Fried2008}. For a recent survey of the efficiency literature, we refer to \cite{Daraio2020}.

First, we build a standard linear regression model, estimate it by the ordinary least squares method and verify whether there is inefficiency present in the data. Technical inefficiency manifests as negatively skewed residuals. All output variables exhibit negative skewness ranging from -0.993 (RQ) to -0.027 (CC). It is therefore suitable to include the inefficiency variable into the model and utilize stochastic frontier analysis. Evenmore, the inefficient component in stochastic frontier models is dominant for all six output variables as its share of the total variance ranges from 0.804 (PV) to 0.922 (VA).

Second, we specify whether efficiencies are static or time-varying. As our main explanatory variables are static, we consider the static model to be more meaningful in our application. To quantify this, we start with the time-varying model of \cite{Battese1992} and find that the trend in efficiencies is quite negligible as its associated coefficient ranges from -0.006 (PV) to 0.003 (GE). We therefore resort to the time-invariant specification in the fashion of \cite{Pitt1981}.

Third, we specify the distribution of the technical inefficiency variable. As proposed by \cite{Stevenson1980}, we start with the two-parameter truncated normal distribution and find that the mode $\mu$ ranges from 0.896 (GE) to 1.089 (CC). Clearly, the parameter $\mu$ is significantly different from 0 and the distribution does not reduce to the half-normal distribution with zero mode. We therefore stick with the truncated normal distribution.

Our final model is described as follows. Let $N$ be the number of countries, $T$ the number of years, $M$ the number of output variables, $K$ the number of input variables, and $L$ the number of control variables. Dependent variable $Y_{itj}$; $i=1,\ldots,N$; $t=1,\ldots,T$; $j=1,\ldots,M$ then follows
$$
Y_{itj} = \alpha_{j} + \sum_{k=1}^K \beta_{jk} X_{itk} + \sum_{l=1}^L \gamma_{jl} Z_{itl} - U_{ij} + V_{itj}, 
$$
where $X_{itk}$ are input variables, $Z_{itl}$ are control variables, $U_{ij}$ are non-negative random variables capturing technical inefficiency, and $V_{itj}$ are random variables representing the error term. We assume that $U_{ij}$ are i.i.d.\ with the normal distribution $\mathrm{N} (\mu_j , \theta_j \sigma_j^2 )$ truncated at zero, $V_{itj}$ are i.i.d.\ with the normal distribution $\mathrm{N} ( 0, (1 - \theta_j) \sigma_j^2 )$, and $U_{ij}$ are independent of $V_{itj}$. Note that we utilize the parametrization of \cite{Battese1977} for $U_{ij}$ and $V_{itj}$. The model includes the constant parameter $\alpha_{j}$, the parameters for the input variables $\beta_{jk}$, the parameters for the control variables $\gamma_{jl}$, the variance of the random component $\sigma_j^2$, the ratio between the variance of the inefficiency variable and the error term $\theta_j$, and the mode of the inefficiency variable $\mu_j$. Using the abbreviations from Section \ref{sec:ind}, we can rewrite the model more specifically as
$$
\begin{aligned}
Y_{itj} &= \alpha_{j} + \beta_{j1} PDI_{it} + \beta_{j2} IDV_{it} + \beta_{j3} MAS_{it} + \beta_{j4} UAI_{it} \\
& \quad + \beta_{j5} LTO_{it} + \beta_{j6} IVR_{it} + \gamma_{j} GDP_{it} - U_{ij} + V_{itj}. \\
\end{aligned}
$$
We estimate the model by the maximum likelihood method.

\section{Results}
\label{sec:res}

\begin{figure}[ht]
\begin{center}
\includegraphics[width=13cm]{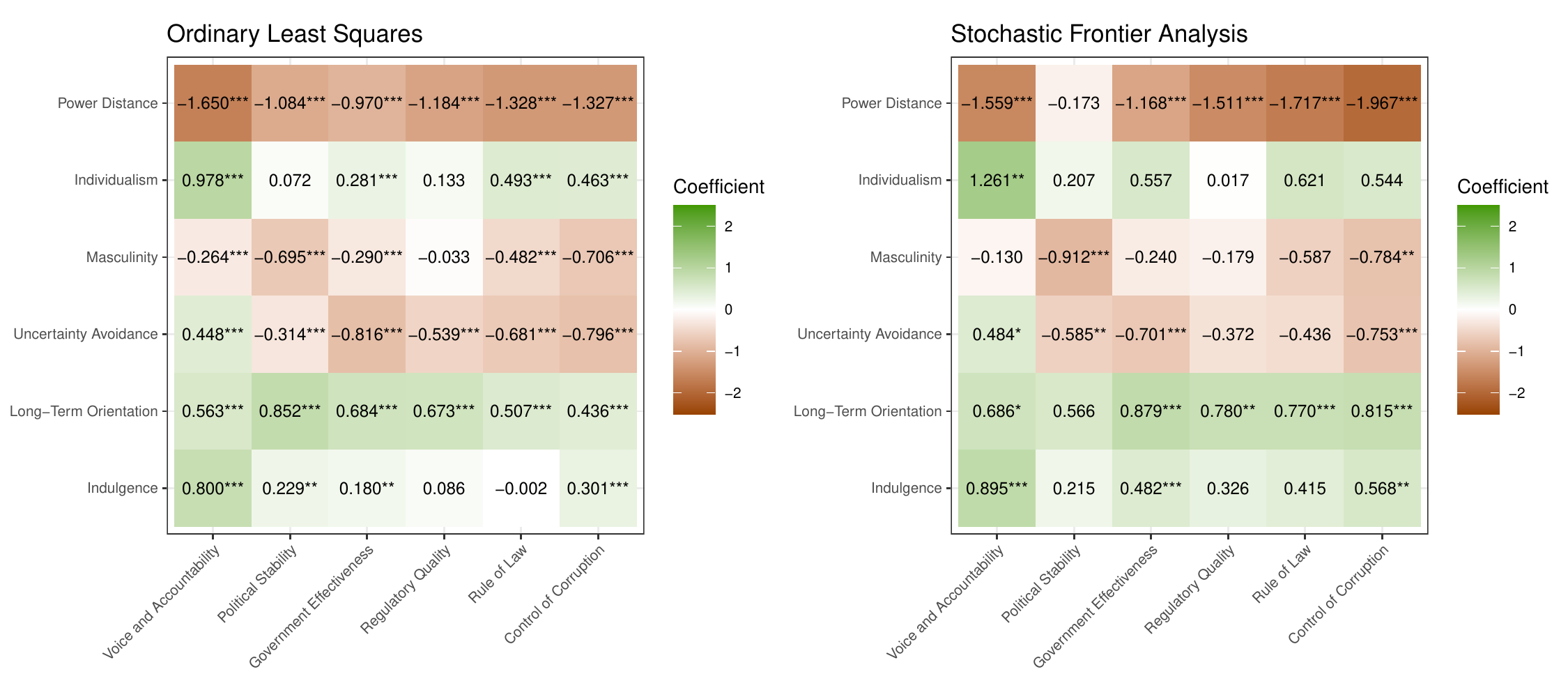}
\caption{Estimated coefficients of the linear regression model and the stochastic frontier model.}
\label{fig:coef}
\end{center}
\end{figure}

\begin{table}[ht]
\begin{center}
\caption{Estimated coefficients with standard errors of the stochastic frontier model.} 
\label{tab:coef} 
\footnotesize
\begin{tabular}{@{\extracolsep{5pt}}lcccccc} 
\hline
& VA & PV & GE & RQ & RL & CC \\
\hline
  Constant & 0.882$^{*}$ & 1.366$^{**}$ & 1.632$^{***}$ & 2.008$^{***}$ & 2.007$^{***}$ & 2.574$^{***}$ \\ 
  & (0.485) & (0.582) & (0.277) & (0.263) & (0.314) & (0.355) \\ 
  & & & & & & \\ 
  Power Distance Index & $-$1.559$^{***}$ & $-$0.173 & $-$1.168$^{***}$ & $-$1.511$^{***}$ & $-$1.717$^{***}$ & $-$1.967$^{***}$ \\ 
  & (0.393) & (0.487) & (0.191) & (0.400) & (0.505) & (0.382) \\ 
  & & & & & & \\ 
  Individualism & 1.261$^{***}$ & 0.207 & 0.557$^{*}$ & 0.017 & 0.621 & 0.544 \\ 
  & (0.458) & (0.502) & (0.287) & (0.374) & (0.531) & (0.347) \\ 
  & & & & & & \\ 
  Masculinity & $-$0.130 & $-$0.912$^{***}$ & $-$0.240 & $-$0.179 & $-$0.587$^{*}$ & $-$0.784$^{***}$ \\ 
  & (0.251) & (0.223) & (0.229) & (0.338) & (0.324) & (0.241) \\ 
  & & & & & & \\ 
  Uncertainty Avoidance & 0.484$^{**}$ & $-$0.585$^{***}$ & $-$0.701$^{***}$ & $-$0.372$^{*}$ & $-$0.436$^{*}$ & $-$0.753$^{***}$ \\ 
  & (0.229) & (0.219) & (0.108) & (0.200) & (0.245) & (0.124) \\ 
  & & & & & & \\ 
  Long-Term Orientation & 0.686$^{**}$ & 0.566$^{*}$ & 0.879$^{***}$ & 0.780$^{***}$ & 0.770$^{***}$ & 0.815$^{***}$ \\ 
  & (0.290) & (0.298) & (0.176) & (0.251) & (0.222) & (0.219) \\ 
  & & & & & & \\ 
  Indulgence & 0.895$^{***}$ & 0.215 & 0.482$^{***}$ & 0.326 & 0.415 & 0.568$^{***}$ \\ 
  & (0.262) & (0.251) & (0.144) & (0.211) & (0.270) & (0.180) \\ 
  & & & & & & \\ 
  GDP Level & 0.076$^{***}$ & 0.397$^{***}$ & 0.318$^{***}$ & 0.346$^{***}$ & 0.281$^{***}$ & 0.264$^{***}$ \\ 
  & (0.016) & (0.026) & (0.017) & (0.018) & (0.016) & (0.016) \\ 
  & & & & & & \\ 
  Parameter $\sigma^2$ & 0.371$^{***}$ & 0.488$^{***}$ & 0.233$^{***}$ & 0.276$^{***}$ & 0.258$^{***}$ & 0.329$^{***}$ \\ 
  & (0.063) & (0.111) & (0.032) & (0.050) & (0.066) & (0.035) \\ 
  & & & & & & \\ 
  Parameter $\theta$ & 0.922$^{***}$ & 0.804$^{***}$ & 0.862$^{***}$ & 0.862$^{***}$ & 0.887$^{***}$ & 0.901$^{***}$ \\ 
  & (0.013) & (0.044) & (0.017) & (0.023) & (0.027) & (0.009) \\ 
  & & & & & & \\ 
  Parameter $\mu$ & 1.065$^{***}$ & 0.912$^{***}$ & 0.896$^{***}$ & 0.976$^{***}$ & 0.958$^{***}$ & 1.089$^{***}$ \\ 
  & (0.115) & (0.213) & (0.101) & (0.090) & (0.319) & (0.074) \\ 
\hline
\multicolumn{7}{r}{\textit{Note:} $^{***}p < 0.001$; $^{**}p < 0.01$; $^{*}p < 0.05$}
\end{tabular} 
\end{center}
\end{table} 

\begin{figure}[ht]
\begin{center}
\includegraphics[width=13cm]{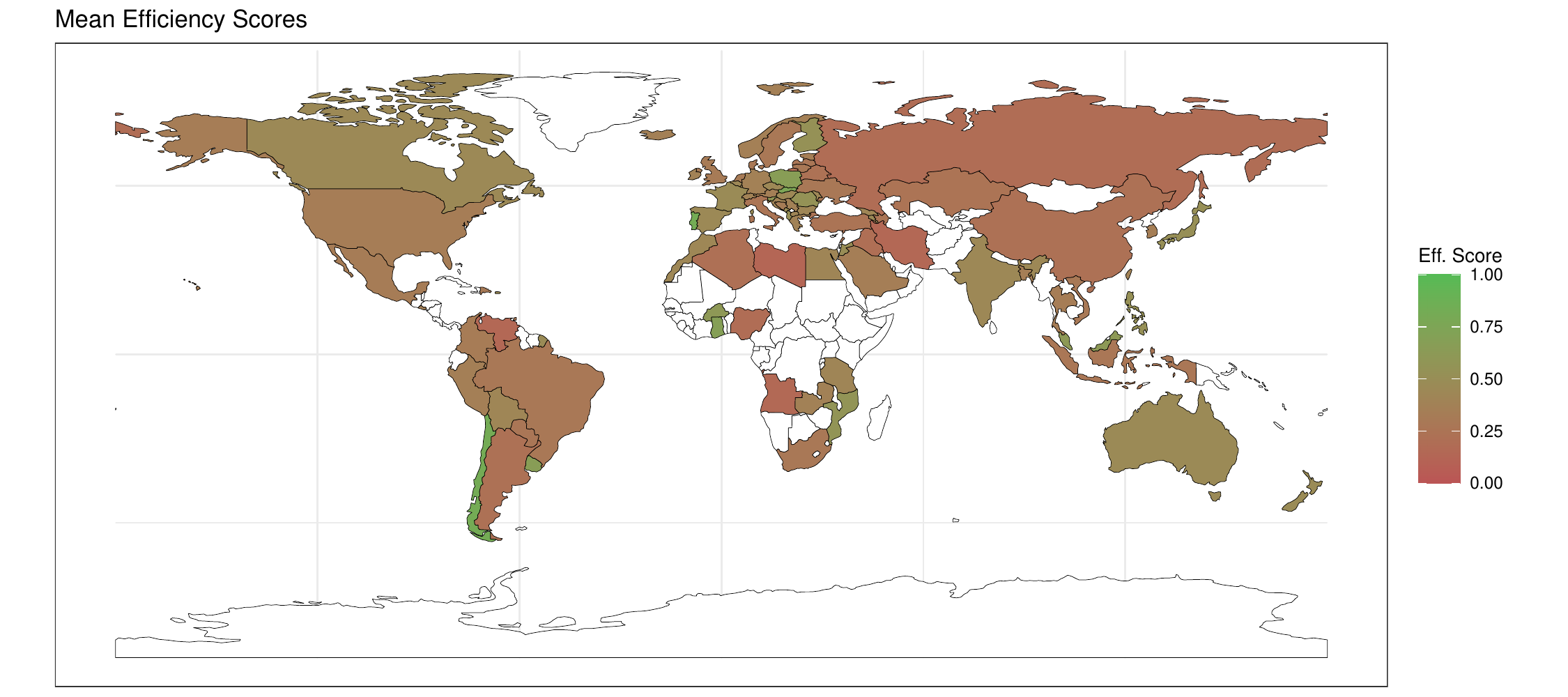}
\caption{Mean efficiency scores obtained from the stochastic frontier model.}
\label{fig:world}
\end{center}
\end{figure}

Our data sample consists of $N=94$ countries observed over $T=21$ time periods from 1996 to 2019 with years 1997, 1999 and 2001 missing. Furthermore, there are additional 13 observations missing. We therefore have 1961 observations in total. The range of countries in our data sample is limited primarily due to the availability of the Hofstede's Cultural Dimensions data, while the range of years is limited due to the availability of the Worldwide Governance Indicators data.

We estimate both the linear regression model and the stochastic frontier model. The estimated coefficients are reported in Figure \ref{fig:coef} and Table \ref{tab:coef}. The models are able to capture a large portion of variability in the output variables. Specifically, the R$^2$ statistic in the regression model ranges from 0.545 (PV) to 0.813 (GE). Both models have very similar values of the coefficients suggesting the robustness of our approach. The significance of the coefficients, however, differs. In the stochastic frontier models, there are much fewer significant variables. As the stochastic frontier model is based on a more general distribution, it is more reliable and we focus solely on it from now on.

We find that the direction of the effect of Hofstede's Cultural Dimensions is consistent across all six Worldwide Governance Indicators. Individualism, Long-Term Orientation, and Indulgence have a positive effect on governance while Power Distance, Masculinity, and Uncertainty Avoidance have a negative effect. The only exception is the effect of Uncertainty Avoidance on Voice and Accountibility, which is positive. Power Distance and Long-Term Orientation play a significant role in five out of six indicators suggesting their universal impact. Masculinity, on the other hand, is found significant only for Political Stability and Control of Corruption; Individualism only for Voice and Accountability. The GDP control variable has a significant positive effect for all indicators. When omitted from the model, however, the results do not distinctly change.

Finally, we examine efficiency of the individual countries. In general, the efficiency score lies in the interval from 0 to 1 with lower values indicating inefficiency. The efficiency averaged over countries ranges from 0.357 (CC) to 0.423 (GE). The efficiency averaged over the output variables is shown in Figure \ref{fig:world} for the individual countries.

The countries that did well in our efficiency analysis are Chile and Uruguay in Latin America, Mozambique, Burkina Faso and Ivory Coast in Africa, Slovakia and Poland and Portugal in Europe as well as Malaysia in Asia. All of these countries have high values in Hofstede's Cultural Dimensions which we identified as a roadblock to good governance and development. All of the above mentioned countries have at least the score of 60 in Power Distance, they are Short-Term Oriented (below 40, with the exception of Slovakia), and generally have a high Uncertainty Avoidance (above 50, with the exception of Malaysia and Mozambique). These countries seem to have good governance relative to their culture.

The other end is represented by countries whose culture would allow for relatively good governance, yet, it is not in place. If we omit countries in civil war (Venezuela and to a lesser extent Argentina in Latin America), these are Angola, Nigeria, and Algeria in Africa, Iran in Asia, and Russia and Belarus in Europe. The rest of the countries fall in the middle of the sample as they seem to have as good governance as their culture would predict.

When sorted by Worldwide Governance Indicators, countries occupied an entire range of possible outcomes in efficiency. A group of three countries namely Portugal, Slovakia, and the former Portuguese colony of Cape Verde ranked highest in the Voice and Accountability indicator, all above 0.9, while four countries namely Iran, Saudi Arabia, Libya and China ranked below 0.1. In the Political Stability and Absence of Violence indicator, São Tomé \& Príncipe, Mozambique, and Slovakia, respectively were the only countries reaching the 0.9 mark, while Iraq, Turkey, and Nigeria, followed closely by Russia were the worst. Again, with the Government Effectiveness indicator only three countries achieved the value of 0.9. These are Malaysia, Portugal, and Chile. Libya, Venezuela, and Angola\footnote{It is of some interest that Portugal and former Portuguese colonies did very well in our results as they have cultures with high Power Distance but decent results in governance indicators. Angola is the only exeption where the Worldwide Governance Indicators are at alarming levels.} were the worst and the only countries of the sample below 0.2 level. Chile and Hong Kong both above 0.9, followed by Slovakia (0.777) were the three best countries in the Regulatory Quality indicator and Libya, Iran, and Venezuela the three worst. Chile, Portugal, and Hong Kong also were the best in Rule of Law while Venezuela, Libya, and Angola were the three worst. Finally, the Control of Corruption indicator saw Singapore, Uruguay, and Chile at the top and Angola, Latvia, and Lithuania at the bottom. The unexpected inclusion of the latter two countries suggests we might see an anti-corruption movement in those countries as the culture of those countries (Power Distance just 44) seems to be hostile to the current levels of corruption.

The overall results in mean efficiency score are as follows: Portugal (0.822), Chile (0.821), Slovakia (0.758), Hong Kong (0.743) and Ghana (0.686) are the top five countries and Libya (0.126), Venezuela (0.143), Iran (0.144), Angola (0.156) and Russia (0.186) are the five bottom ones.

\section{Conclusion}
\label{sec:con}

We have found an albeit rudimentary measurement of the strength of the relationship between cultural characteristics of countries and the quality of their institutions in 94 countries over 1996--2019. We have also determined efficiency scores for the transformation of cultural characteristics into institutions in each of those individual countries.

The results of our study pose several policy implications and avenues for further research. After more detailed analysis, policies on foreign aid levels can be improved as it might be possible to discern which countries' governance can be relatively successfully improved and/or which countries' governments are more effective in reforming national institutions for example in situations of adverse cultural patterns. The migration policies of countries deciding on levels and source countries of economic migration can also be improved as the difference between the influence of culture and governance in each country becomes clearer. There are also significant implications for international security studies and related fields.

There are several possible directions for the development of our model. In the current paper, we have considered GDP per capita as a control variable but did not further investigate its role in countries' culture and governance. As this is a hot topic in the literature, it will certainly be interesting to study this relationship in more detail. Migration and heterogeneity of the culture within a country could also be included in the model. Finally, we could improve our model by adding a spatial structure to capture dependency between countries.

\section*{Funding}
\label{sec:fund}

The work of Vladimír Holý was supported by the Czech Science Foundation under project 19-08985S.

%\bibliography{library.bib,data.bib}
%\bibliographystyle{mynatstyle}

\end{document}